\documentclass{article}
\usepackage[utf8x]{inputenc}
\usepackage{spconf,amsmath,graphicx}

\graphicspath{{figures/}}

\usepackage{cite}
\usepackage{amsmath,amssymb,amsfonts}
\usepackage{algorithmic}
\usepackage{graphicx}
\usepackage{multirow}
\usepackage{tikz}
\usepackage{hyperref}
\usepackage{xcolor}
\usepackage{lipsum}
\usepackage{caption}
\usepackage{siunitx}
\usepackage{commath}
\usepackage{subfig}
\usepackage{pgfplots}
\pgfplotsset{compat=1.17}
\usepackage{mathrsfs}
\usetikzlibrary{arrows}
\usetikzlibrary{positioning}
\usetikzlibrary{3d}
\usepackage{textcomp}
\usepackage{xcolor}
\def\BibTeX{{\rm B\kern-.05em{\sc i\kern-.025em b}\kern-.08em
    T\kern-.1667em\lower.7ex\hbox{E}\kern-.125emX}}
 
\usepackage{hyperref}
\hypersetup{
	colorlinks=true,
	linkcolor=blue,
	filecolor=magenta,      
	urlcolor=cyan,
	pdftitle={Self-Organized Variational Autoencoders for Learned Image Compression},
	pdfauthor={M. A. Yılmaz; O. Keleş; H. Güven; M. Tekalp},
	pdfstartview=Fit,
	pdffitwindow=True,
}

\begin{document}
\definecolor{col_w}{rgb}{0.870588,0.796078,0.776470}
\definecolor{col_wn}{rgb}{0.996078,0.847059,0.364706}
\definecolor{col_out}{rgb}{1,0.5,0}
\definecolor{col_bc}{rgb}{0,0.5,1}
\definecolor{col_conv}{rgb}{0,1,1}
\definecolor{col_out}{rgb}{1,0.5,0}
\title{Self-Organized Variational Autoencoders (Self-VAE)  \\ for Learned Image Compression}

\name{M. Akın Yılmaz$^1$, Onur Keleş$^1$, Hilal Güven$^1$, A. Murat Tekalp$^1$, Junaid Malik$^2$, Serkan Kıranyaz$^3$
\thanks{This work was supported by TUBITAK projects 217E033 and 120C156, and a grant from Turkish Is Bank to KUIS AI Center. A. M. Tekalp also acknowledges support from Turkish Academy of Sciences (TUBA).}}

\address{$^1$Dept. of Electrical \& Electronics Eng., Koç University, 34450 İstanbul, Turkey \\
$^2$Tampere University, Tampere, Finland \\
$^3$Qatar University, Doha, Qatar }

\maketitle

\begin{abstract}
In end-to-end optimized learned image compression, it is standard practice to use a convolutional variational autoencoder with generalized divisive normalization (GDN) to transform images into a latent space. Recently, Operational Neural Networks (ONNs) that learn the best non-linearity from a set of alternatives, and their “self-organized” variants, Self-ONNs, that approximate any non-linearity via Taylor series have been proposed to address the limitations of convolutional layers and a fixed nonlinear activation. In this paper, we propose to replace the convolutional and GDN layers in the variational autoencoder with self-organized operational layers, and propose a novel self-organized variational autoencoder (Self-VAE) architecture that benefits from stronger non-linearity. The experimental results demonstrate that the proposed Self-VAE yields improvements in both rate-distortion performance and perceptual image quality.
\end{abstract}

\vspace{2pt}

\section{Introduction}
\label{intro}


End-to-end optimization of rate-distortion (RD) performance using nonlinear transforms via a variational autoencoder with generalized divisive normalization (GDN) nonlinearity~\cite{gdn} was first proposed in \cite{balle_pcs,balle_end} and \cite{compressive_ae}.
In recent years, improvements in the learned entropy models that is an integral part of this framework enabled learned image compression methods  \cite{balle_scale,minnen_joint,variational_low,cheng2020image,minnen2020channelwise} to outperform standard image compression methods such as JPEG \cite{jpeg}, JPEG2000 \cite{jpeg2000}, and BPG~\cite{bpg}. 

Independent from recent developments in learned image compression, Operational Neural Networks (ONNs)~\cite{kiranyaz2020operational,kiranyaz2017progressive,kiranyaz2017generalized,thanh2018progressive,tran2019heterogeneous,tran2019knowledge} that are heterogeneous networks with nonlinear nodal operators, and their “self-organized” variants, Self-ONNs~\cite{kiranyaz2020self,malik2020}, with a novel generative neuron model that can approximate the nodal operators via Taylor series have been proposed as more powerful network models compared to convolutional networks, which have homogeneous configuration.

In this paper, based on these recent works, we propose a self-organized variational autoencoder (Self-VAE) by replacing the convolutional and GDN layers in the generic variational autoencoder with the operational layers of Self-ONNs in order to benefit from stronger non-linearity. The~rest of the paper is organized as follows: Section~\ref{related} discusses related works and our contributions. The proposed self-organized variational autoencoder architecture is introduced in Section~\ref{selfvae}. Section~\ref{eval} presents experimental results for image compression. Finally, Section~\ref{conc} concludes the paper.

\begin{figure}
    \centering
    \begin{tabular}{ccc}
        \multirow{2}{*}{\raisebox{-\height}{\begin{tikzpicture}
            \node[anchor=south west,inner sep=1] (img) at (0,0) {\includegraphics[width=0.48\columnwidth]{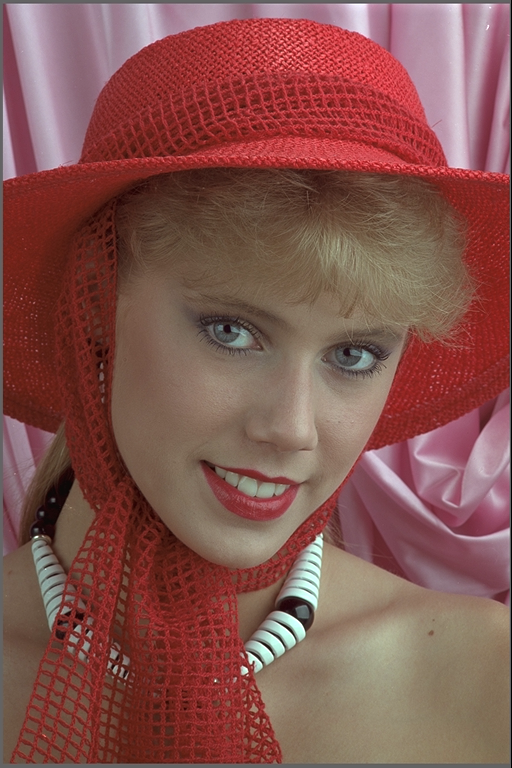}}; 
		    \begin{scope}[x={(img.south east)},y={(img.north west)}]
		    	\draw[green,thick] (0.5664,0.7266) rectangle (0.7813,0.4922);
		    \end{scope}  
	\end{tikzpicture}}} & \raisebox{-\height}{\begin{minipage}{0.20\columnwidth}
        \includegraphics[width=\columnwidth]{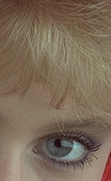}\vspace{-10pt}
        \caption*{Original}
        \label{fig:kdm04_org}
    \end{minipage}} & \raisebox{-\height}{\begin{minipage}{0.20\columnwidth}
        \includegraphics[width=\columnwidth]{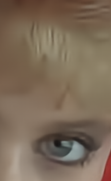}\vspace{-10pt}
        \caption*{Self-VAE}
        \label{fig:kdm04_selfonn}
    \end{minipage}} \\
         & \begin{minipage}{0.20\columnwidth}
        \includegraphics[width=\columnwidth]{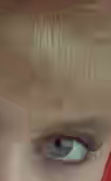}\vspace{-10pt}
        \caption*{BPG}
        \label{fig:kdm04_bpg}
    \end{minipage} & \begin{minipage}{0.20\columnwidth}
        \includegraphics[width=\columnwidth]{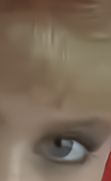}\vspace{-10pt}
        \caption*{GDN~\cite{minnen_joint}}
        \label{fig:kdm04_minnen}
    \end{minipage}
    \end{tabular}
    \caption{Visual evaluation of original vs. reconstructed crops from \textit{Kodim04} in Kodak dataset at approximately 0.1 bpp. The hair is sharper in Self-VAE (same entropy model as~\cite{minnen_joint}).}
    \label{fig:my_label}
\end{figure}

\section{Related work and Contributions}
\label{related} \vspace{-4pt}

\subsection{\textbf{Image Compression}} \vspace{-2pt}
The state of the art in image compression by nonlinear transform coding is summarized in~\cite{nonlinear}. The improvements over time were achieved by better entropy modeling. In \cite{balle_scale,minnen_joint,variational_low,cheng2020image,minnen2020channelwise},  entropy models with different complexity were used to model the distribution of latent variables to achieve better compression performance. In \cite{balle_scale,minnen_joint}, the latent~distribution is modelled by a Gaussian, whose mean and/or scale are estimated by an additional hyperprior network. In \cite{minnen_joint}, this is combined with an autoregressive model to jointly optimize latent distribution parameters. In \cite{cheng2020image}, a Gaussian mixture model is used as a more flexible entropy model and attention modules are added to minimize the distortion between original and reconstructed images. 
Sequential channel-conditional slices and latent residual prediction are proposed in \cite{minnen2020channelwise}.
 
In all these works, GDN layers, which are expected to provide Gaussian latent variables \cite{gdn}, are the only source of nonlinearity in the variational autoencoder to achieve nonlinear transform coding. In this work, we replace convolutional and GDN layers with self-organized operational layers to achieve stronger nonlinearity and better performance.

\subsection{\textbf{Self-ONNs}} \vspace{-2pt}
ONNs based on Generalized Operational Perceptrons \cite{kiranyaz2020operational,kiranyaz2017progressive,kiranyaz2017generalized,thanh2018progressive,tran2019heterogeneous,tran2019knowledge}, and their self-organized variants, Self-ONNs \cite{kiranyaz2020self,malik2020}, based on the generative neuron model (see Section 3.1) have recently been proposed as heterogeneous network models that are capable of performing any non-linear kernel transformation. While conventional ONNs have to search for the best kernel operators from a pre-determined operator set library, Self-ONNs can generate the kernel operator by a Taylor series approximation of the desired order. It has been noted that exhaustive search for the best transformation makes ONNs computationally expensive~\cite{kiranyaz2020operational,kiranyaz2020self,malik2020b}; hence, Self-ONNs have become a more promising and efficient network architecture that can provide an utmost level of diversity. In this work, we extend Self-ONNs to Self-VAEs in order to use them for end-to-end RD-optimized nonlinear transform image compression for the first time.

\subsection{\textbf{Contributions}}
The main contributions of this paper are: \vspace{-5pt}
\begin{itemize} 
\item We propose a novel self-organized variational autoencoder (Self-VAE) architecture by replacing the convolutional and GDN layers in a convolutional VAE with self-organized operational layers (SOL) in Section~\ref{selfvae} \vspace{-6pt}
\item We employ the Self-VAE as encoder and decoder networks in the nonlinear transform coding paradigm \cite{minnen_joint}, benefiting from the ability of SOLs to learn highly complex functions \cite{kiranyaz2020self} in Section~\ref{eval}.
\end{itemize} \vspace{-3pt}


\section{Self-VAE for Image Compression}
\label{selfvae}
We first present the concept of self-organized operational layers (SOL) in Section 3.1. We then introduce the Self-VAE architecture for the image compression task in Section 3.2. 

\subsection{Self-Organized Operational Layer (SOL)}

\begin{figure}
		\centering
		\resizebox{0.485\textwidth}{!}
		{\begin{tikzpicture}[line join=round,>=triangle 45,x=0.75cm,y=0.75cm]
				\node[canvas is yz plane at x=0] (temp) at (0,0,0) {\includegraphics[width=4cm,height=4cm]{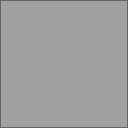}};
				\node[canvas is yz plane at x=0] (temp) at (0.25,0,0) {\includegraphics[width=4cm,height=4cm]{tensor_img.png}};
				\node[canvas is yz plane at x=0] (temp) at (0.5,0,0) {\includegraphics[width=4cm,height=4cm]{tensor_img.png}};
				
				\node[canvas is yz plane at x=0] (temp) at (2,0,0) {\includegraphics[width=4cm,height=4cm]{tensor_img.png}};
				
				\node[canvas is yz plane at x=0] (temp) at (32,0,0) {\includegraphics[width=4cm,height=4cm]{tensor_img.png}};
				
				\draw [line width=2pt] (6.5,1.5) -- (9.5,1.5) -- (9.5,-1.5) -- (6.5,-1.5) -- cycle;
				\draw [line width=2pt] (6.5,-5.5) -- (9.5,-5.5) -- (9.5,-8.5) -- (6.5,-8.5) -- cycle;
				
				
				\fill[line width=0.4pt,color=col_wn,fill=col_wn,fill opacity=0.75] (12,-4) -- (18,-4) -- (18,-10) -- (12,-10) -- cycle;
				
				\fill[line width=0.4pt,color=col_wn,fill=col_wn,fill opacity=0.75] (12,3) -- (18,3) -- (18,-3) -- (12,-3) -- cycle;
				
				\fill[line width=0.4pt,color=col_wn,fill=col_wn,fill opacity=0.75] (12,10) -- (18,10) -- (18,4) -- (12,4) -- cycle;
				
				\fill[line width=0.4pt,color=blue,fill=col_bc] (14.5,9.5) -- (17.5,9.5) -- (17.5,6.5) -- (14.5,6.5) -- cycle;
				\draw [line width=0.75pt,dash pattern=on 4pt off 4pt] (16.5,9.5)-- (16.5,6.5);
				\draw [line width=0.75pt,dash pattern=on 4pt off 4pt] (15.5,9.5)-- (15.5,6.5);
				\draw [line width=0.75pt,dash pattern=on 4pt off 4pt] (14.5,8.5)-- (17.5,8.5);
				\draw [line width=0.75pt,dash pattern=on 4pt off 4pt] (14.5,7.5)-- (17.5,7.5);
				
				\fill[line width=0.4pt,color=green,fill=green] (13,8) -- (16,8) -- (16,5) -- (13,5) -- cycle;
				\draw [line width=0.75pt,dash pattern=on 4pt off 4pt] (15,8)-- (15,5);
				\draw [line width=0.75pt,dash pattern=on 4pt off 4pt] (14,8)-- (14,5);
				\draw [line width=0.75pt,dash pattern=on 4pt off 4pt] (13,7)-- (16,7);
				\draw [line width=0.75pt,dash pattern=on 4pt off 4pt] (13,6)-- (16,6);
				
				\fill[line width=0.4pt,color=red,fill=red] (12.5,7.5) -- (15.5,7.5) -- (15.5,4.5) -- (12.5,4.5) -- cycle;
				\draw [line width=0.75pt,dash pattern=on 4pt off 4pt] (14.5,7.5)-- (14.5,4.5);
				\draw [line width=0.75pt,dash pattern=on 4pt off 4pt] (13.5,7.5)-- (13.5,4.5);
				\draw [line width=0.75pt,dash pattern=on 4pt off 4pt] (12.5,6.5)-- (15.5,6.5);
				\draw [line width=0.75pt,dash pattern=on 4pt off 4pt] (12.5,5.5)-- (15.5,5.5);
				
				\fill[line width=0.4pt,color=blue,fill=col_bc] (14.5,2.5) -- (17.5,2.5) -- (17.5,-0.5) -- (14.5,-0.5) -- cycle;
				\draw [line width=0.75pt,dash pattern=on 4pt off 4pt] (16.5,2.5)-- (16.5,-0.5);
				\draw [line width=0.75pt,dash pattern=on 4pt off 4pt] (15.5,2.5)-- (15.5,-0.5);
				\draw [line width=0.75pt,dash pattern=on 4pt off 4pt] (14.5,1.5)-- (17.5,1.5);
				\draw [line width=0.75pt,dash pattern=on 4pt off 4pt] (14.5,0.5)-- (17.5,0.5);
				
				\fill[line width=0.4pt,color=green,fill=green] (13,1) -- (16,1) -- (16,-2) -- (13,-2) -- cycle;
				\draw [line width=0.75pt,dash pattern=on 4pt off 4pt] (15,1)-- (15,-2);
				\draw [line width=0.75pt,dash pattern=on 4pt off 4pt] (14,1)-- (14,-2);
				\draw [line width=0.75pt,dash pattern=on 4pt off 4pt] (13,0)-- (16,0);
				\draw [line width=0.75pt,dash pattern=on 4pt off 4pt] (13,-1)-- (16,-1);
				
				\fill[line width=0.4pt,color=red,fill=red] (12.5,0.5) -- (15.5,0.5) -- (15.5,-2.5) -- (12.5,-2.5) -- cycle;
				\draw [line width=0.75pt,dash pattern=on 4pt off 4pt] (14.5,0.5)-- (14.5,-2.5);
				\draw [line width=0.75pt,dash pattern=on 4pt off 4pt] (13.5,0.5)-- (13.5,-2.5);
				\draw [line width=0.75pt,dash pattern=on 4pt off 4pt] (12.5,-0.5)-- (15.5,-0.5);
				\draw [line width=0.75pt,dash pattern=on 4pt off 4pt] (12.5,-1.5)-- (15.5,-1.5);
				
				\fill[line width=0.4pt,color=blue,fill=col_bc] (14.5,-4.5) -- (17.5,-4.5) -- (17.5,-7.5) -- (14.5,-7.5) -- cycle;
				\draw [line width=0.75pt,dash pattern=on 4pt off 4pt] (16.5,-4.5)-- (16.5,-7.5);
				\draw [line width=0.75pt,dash pattern=on 4pt off 4pt] (15.5,-4.5)-- (15.5,-7.5);
				\draw [line width=0.75pt,dash pattern=on 4pt off 4pt] (14.5,-5.5)-- (17.5,-5.5);
				\draw [line width=0.75pt,dash pattern=on 4pt off 4pt] (14.5,-6.5)-- (17.5,-6.5);
				
				\fill[line width=0.4pt,color=green,fill=green] (13,-6) -- (16,-6) -- (16,-9) -- (13,-9) -- cycle;
				\draw [line width=0.75pt,dash pattern=on 4pt off 4pt] (15,-6)-- (15,-9);
				\draw [line width=0.75pt,dash pattern=on 4pt off 4pt] (14,-6)-- (14,-9);
				\draw [line width=0.75pt,dash pattern=on 4pt off 4pt] (13,-7)-- (16,-7);
				\draw [line width=0.75pt,dash pattern=on 4pt off 4pt] (13,-8)-- (16,-8);
				
				\fill[line width=0.4pt,color=red,fill=red] (12.5,-6.5) -- (15.5,-6.5) -- (15.5,-9.5) -- (12.5,-9.5) -- cycle;
				\draw [line width=0.75pt,dash pattern=on 4pt off 4pt] (14.5,-6.5)-- (14.5,-9.5);
				\draw [line width=0.75pt,dash pattern=on 4pt off 4pt] (13.5,-6.5)-- (13.5,-9.5);
				\draw [line width=0.75pt,dash pattern=on 4pt off 4pt] (12.5,-7.5)-- (15.5,-7.5);
				\draw [line width=0.75pt,dash pattern=on 4pt off 4pt] (12.5,-8.5)-- (15.5,-8.5);
				
				\draw [line width=2pt] (24,0) circle (1.5cm);
				
				\draw [->,line width=1pt] (2,0) -- (6.5,0);
				
				\draw [->,line width=1pt] (9.5,0) -- (12,0);
				\draw [line cap=round,line width=2pt] (5,0)-- (5,7);
				\draw [->,line width=1pt] (5,7) -- (12,7);
				\draw [line cap=round,line width=2pt] (5,0)-- (5,-7);
				\draw [->,line width=1pt] (5,-7) -- (6.5,-7);
				\draw [->,line width=1pt] (9.5,-7) -- (12,-7);
				\begin{scriptsize}
					\draw [fill=black] (5,0) circle (3pt);
				\end{scriptsize}
				
				\draw [->,line width=1pt] (18,7) -- (22.7,1.45);
				\draw [->,line width=1pt] (18,0) -- (22,0);
				\draw [->,line width=1pt] (18,-7) -- (22.7,-1.45);
				\draw [->,line width=1pt] (24,-7) -- (24,-2);
				
				\draw [->,line width=1pt] (26,0) -- (31,0);
				
				\begin{scriptsize}
					\draw [fill=black] (0.25,0) circle (1.5pt);
					\draw [fill=black] (0.5,0) circle (1.5pt);
					\draw [fill=black] (0.75,0) circle (1.5pt);
					
					\draw [fill=black] (16.5,5.5) circle (1.5pt);
					\draw [fill=black] (16.75,5.75) circle (1.5pt);
					\draw [fill=black] (17,6) circle (1.5pt);
					
					\draw [fill=black] (16.5,-1.5) circle (1.5pt);
					\draw [fill=black] (16.75,-1.25) circle (1.5pt);
					\draw [fill=black] (17,-1) circle (1.5pt);
					
					\draw [fill=black] (16.5,-8.5) circle (1.5pt);
					\draw [fill=black] (16.75,-8.25) circle (1.5pt);
					\draw [fill=black] (17,-8) circle (1.5pt);
				\end{scriptsize}
				
				\draw [line width=4pt,dash pattern=on 8pt off 8pt] (4,11) -- (29,11) -- (29,-11) -- (4,-11) -- cycle;
				
				\draw (8,0) node[anchor=center] {\Huge $ \mathbf{(\cdot)^{2}} $};
				\draw (8,-7) node[anchor=center] {\Huge $ \mathbf{(\cdot)^{3}} $};
				
				\draw (17.25,-9.5) node[anchor=center] {\Huge $ w_{3} $};
				\draw (17.25,-2.5) node[anchor=center] {\Huge $ w_{2} $};
				\draw (17.25,4.5) node[anchor=center] {\Huge $ w_{1} $};
				\draw (24,-7.75) node[anchor=center] {\Huge $ w_{0} $};
				
				\draw (24,0) node[anchor=center] {\Huge $ \mathbf{+} $};
				
				\draw (1,-4) node[anchor=center] {\Huge Input};
				\draw (1,-5) node[anchor=center] {\Huge Tensor};
				
				\draw (26,10) node[anchor=center] {\Huge Generative};
				\draw (26,9) node[anchor=center] {\Huge Neuron};
				
				\draw (32,-4) node[anchor=center] {\Huge Channel};
				
			\end{tikzpicture}
		}
		\caption{Illustration of a generative neuron for $q=3$.}
		\label{fig:gen_nrn}
	\end{figure}

Assume that a desired kernel operator transformation $f(x)$ has a Taylor series expansion
\begin{equation}
    \label{eq:TaylorSeries_f}
    f(x) = \sum_{n=0}^{\infty}\dfrac{f^{(n)}(a)}{n!}(x-a)^{n}
\end{equation}
around the point $ a $. If we truncate the series to $q$ terms, we~have the approximation $g(\mathbf{w},x,a)$ given by
\begin{align}
	\label{eq:truncdTaylor}
	g(\mathbf{w},x,a) &= w_{0} + w_{1}(x-a) + \cdots + w_{q}(x-a)^{q}
\end{align}
where
\begin{equation}
	\label{eq:weights}
	w_{n} = \dfrac{f^{(n)}(a)}{n!}.
\end{equation}
A generative neuron with $3\times3$ kernels, $a=0$ and $q=3$ is illustrated in Figure \ref{fig:gen_nrn}. Each~neuron takes a tensor as input and outputs a single channel. The outputs must be limited within a small range around $a$ before they are input to the~next generative neuron, which will also perform a Taylor series expansion around $a$. Setting~$a=0$, the activation can be taken as $\tanh(x)$ to bound the outputs in the~range~$\left[-1\,\,1\right]$. Note that if we choose $q=1$ and $a=0$, the generative neuron reduces to the classic convolutional neuron.

A SOL is formed by $c_{out}$ generative neurons. For a $c_{in}$-channel input tensor, a SOL is composed of $c_{out}$ banks of $q$-branch filterbanks, where each filterbank represents a generative neuron, depicted in Figure~\ref{fig:gen_nrn}, with the parameters $w_n$, $n=1,2,\dots,q$ denoting $c_{in}$-channel convolution kernels and $w_0$ is a bias term. These parameters can be learned during training using the back-propagation algorithm.

\subsection{Self-Organized Variational AutoEncoder (Self-VAE)}
\begin{figure}[t]
\centering
	\includegraphics[width=0.50\textwidth]{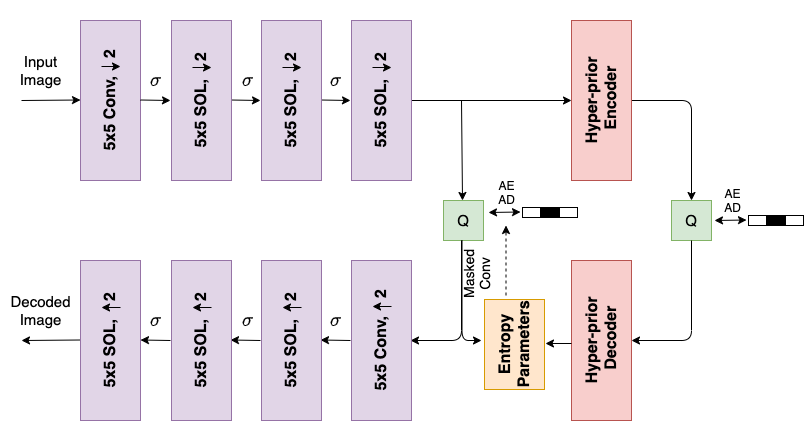} \vspace{-14pt} \\
\caption{Self-VAE network with 4-SOL (purple boxes) for image compression. $\sigma$ is set to tanh to scale output of each layer. Q, AE, AD denote quantization, arithmetic encoding and decoding, respectively. Hyper-prior nets are the same as in~\cite{minnen_joint} }
\label{fig:onn_net}
\end{figure}

While any convolutional VAE can be converted into a Self-VAE, in this work, we have taken the VAE architecture in~\cite{minnen_joint} as the baseline model. We replace convolution layers followed by GDN activations in the encoder and decoder by SOL with $q=3$ assigned to each generative neuron. We use $5\times 5$ kernels with stride~2. $\tanh(x)$ is used to scale inputs to each SOL between -1 and 1, since each SOL implements Taylor series expansion about~0.  The detailed network architecture is shown in Figure~\ref{fig:onn_net}. 

We use the same quantization modeling, hyper-prior encoder/decoder, masked convolutions for context modeling, entropy parameters, and arithmetic encoder/decoder as in~\cite{minnen_joint} for a fair comparison of the effect of the proposed Self-VAE.

\section{Evaluation}
\label{eval}
We evaluate the performance of the Self-VAE in the image compression task.
The details of the training process are discussed in~Section 4.1. Experimental setup and results are presented in Section 4.2 and 4.3, respectively.

\subsection{Training Details}
We train our self-VAE image compression network, where each SOL has $c=192$ channels, on the Vimeo-90k dataset~\cite{vimeo}. In order to construct the RD curve, we trained the network for four different \textit{$\lambda=0.012, 0.026, 0.042, 0.056$}. During training, we apply $256 \times 256$ random cropping for data augmentation. The~mini-batch size is set to 16 and optimization is performed for 1M steps. Adam optimizer\cite{adam} is used for the training process. The initial learning rate is set to 1e-4 and dropped to 1e-5 at 500K steps. We also apply gradient clipping on the norm of the gradients with maximum norm value is set to 1.

\subsection{Experimental Setup}
We evaluate the performance of Self-VAE on the commonly used Kodak images, which consists of 24 uncompressed images. To evaluate the rate-distortion performance we used peak signal-to-noise ration (PSNR) for the quality metric, and bits-per-pixel is used for the rate of compression. 

\begin{figure}[t]
\centering
	\includegraphics[scale=0.35]{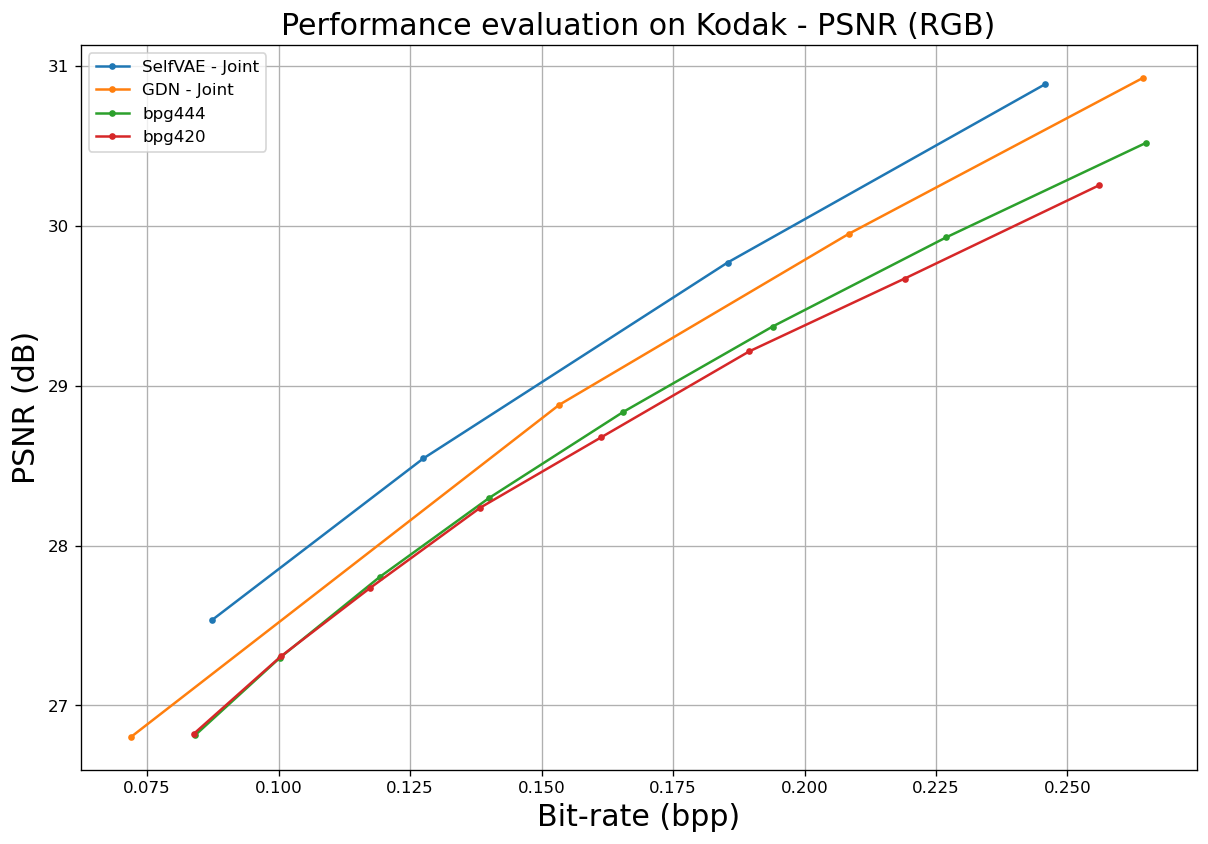} \vspace{-14pt} \\
\caption{Comparison of rate-distortion (RD) performance of Self-VAE vs. GDN layer \cite{minnen_joint} and BPG over KODAK images.}
\label{fig:rd_curve}
\end{figure}

\begin{figure}[t]
\centering
	\includegraphics[scale=0.30]{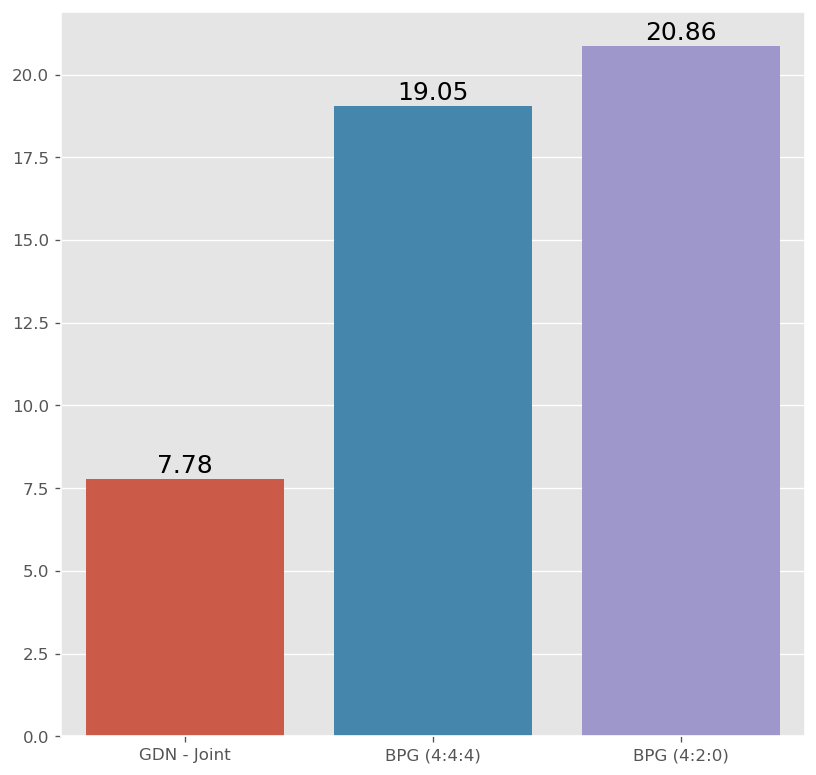} \vspace{-6pt} \\
\caption{Average percent BD-rate improvements (RGB PSNR) for Self-VAE vs.~GDN layer~using~the~same~joint entropy~model~\cite{minnen_joint} and BPG over KODAK images.}
\label{fig:rate_savings}
\end{figure}

\begin{table*}[ht]
\centering
\caption{PSNR and LPIPS values for selected KODAK images. Best results are shown in bold.} \vspace{-8pt}
\resizebox{0.98\textwidth}{!}{
\begin{tabular}{|c|c|c|c|c|c|c|c|c|c|} 
\hline
\multirow{2}{*}{} & \multicolumn{3}{c|}{KODIM 04} & \multicolumn{3}{c|}{KODIM 14} & \multicolumn{3}{c|}{KODIM 15}  \\ \cline{2-10}
       & BPG    & GDN-Joint\cite{minnen_joint} & Ours   & BPG    & GDN-Joint\cite{minnen_joint} & Ours   & BPG    & GDN-Joint\cite{minnen_joint} & Ours   \\ \hline
PSNR       & 30.03 & 30.51 & {\bf 30.69} & 27.52 & 27.92 & {\bf 28.07} & 30.34 & 30.62 & {\bf 30.79}  \\ \hline
LPIPS-Alex & 0.3413 & 0.3565 & {\bf 0.3274} & 0.3459 & 0.3689 & {\bf 0.3361} & 0.2762 & 0.2955 & {\bf 0.2636} \\ \hline
LPIPS-VGG  & 0.4075 & 0.3976 & {\bf 0.3830} & 0.3923 & 0.3922 & {\bf 0.3654} & 0.4083 & 0.4013 & {\bf 0.3770} \\ \hline
\end{tabular}
}
\label{lpips}
\end{table*}
\begin{figure*}[t!]
    \centering
    \begin{minipage}{0.475\columnwidth}
        \caption*{Original}\vspace{-10pt}
        \begin{tikzpicture}
		    \node[anchor=south west,inner sep=1] (img) at (0,0) {\includegraphics[width=\columnwidth]{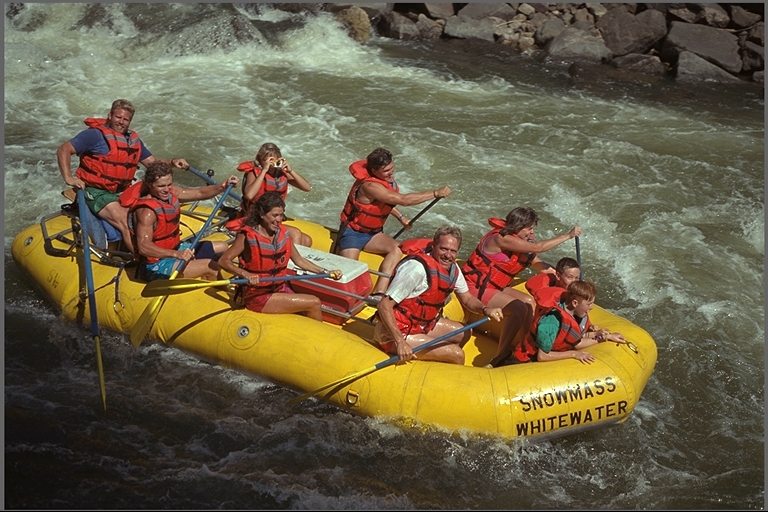}}; 
		    \begin{scope}[x={(img.south east)},y={(img.north west)}]
		    	\draw[green,thick] (0.1042,0.8242) rectangle (0.2018,0.7266);
		    \end{scope}  
	    \end{tikzpicture}
        \caption*{BPP / PSNR}\vspace{-3pt}
        \label{fig:kdm14_org}
    \end{minipage}
    \begin{minipage}{0.475\columnwidth}
        \caption*{BPG}\vspace{-10pt}
        \begin{tikzpicture}
		    \node[anchor=south west,inner sep=1] (img) at (0,0) {\includegraphics[width=\columnwidth]{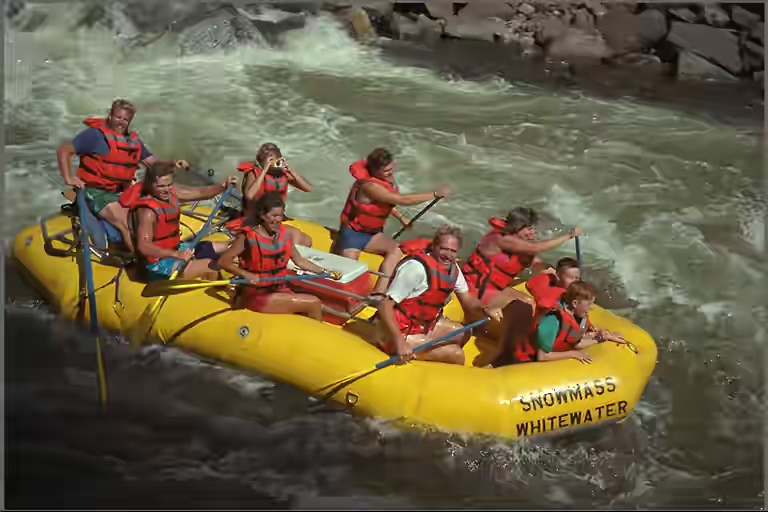}}; 
		    \begin{scope}[x={(img.south east)},y={(img.north west)}]
		    	\draw[green,thick] (0.1042,0.8242) rectangle (0.2018,0.7266);
		    \end{scope}  
	    \end{tikzpicture}
        \caption*{0.258 / 27.52}\vspace{-3pt}
        \label{fig:kdm14_bgg}
    \end{minipage}
    \begin{minipage}{0.475\columnwidth}
        \caption*{GDN-Joint\cite{minnen_joint}}\vspace{-10pt}
        \begin{tikzpicture}
		    \node[anchor=south west,inner sep=1] (img) at (0,0) {\includegraphics[width=\columnwidth]{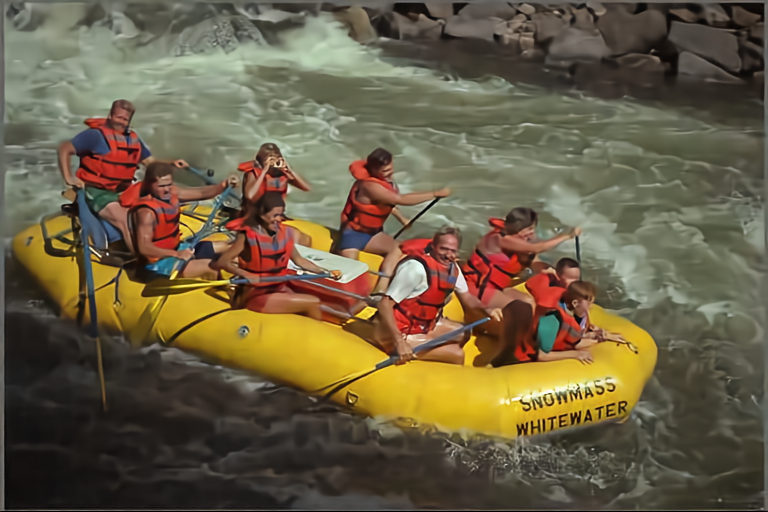}}; 
		    \begin{scope}[x={(img.south east)},y={(img.north west)}]
		    	\draw[green,thick] (0.1042,0.8242) rectangle (0.2018,0.7266);
		    \end{scope}  
	    \end{tikzpicture}
        \caption*{0.257 / 27.92}\vspace{-3pt}
        \label{fig:kdm14_joint}
    \end{minipage}
    \begin{minipage}{0.475\columnwidth}
        \caption*{SelfVAE-Joint}\vspace{-10pt}
        \begin{tikzpicture}
		    \node[anchor=south west,inner sep=1] (img) at (0,0) {\includegraphics[width=\columnwidth]{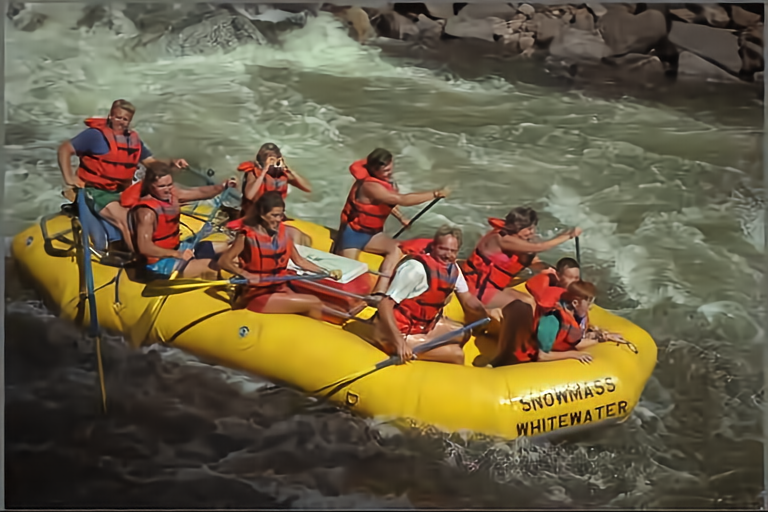}}; 
		    \begin{scope}[x={(img.south east)},y={(img.north west)}]
		    	\draw[green,thick] (0.1042,0.8242) rectangle (0.2018,0.7266);
		    \end{scope}  
	    \end{tikzpicture}
        \caption*{0.256 / 28.07}\vspace{-3pt}
        \label{fig:kdm14_selfonn}
    \end{minipage}\\
    \begin{minipage}{0.475\columnwidth}
        \includegraphics[width=\columnwidth]{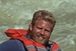}
        \label{fig:kdm14_org}
    \end{minipage}
    \begin{minipage}{0.475\columnwidth}
        \includegraphics[width=\columnwidth]{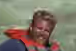}
        \label{fig:kdm14_org}
    \end{minipage}
    \begin{minipage}{0.475\columnwidth}
        \includegraphics[width=\columnwidth]{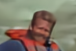}
        \label{fig:kdm14_org}
    \end{minipage}
    \begin{minipage}{0.475\columnwidth}
        \includegraphics[width=\columnwidth]{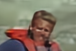}
        \label{fig:kdm14_org}
    \end{minipage}\\
    \vspace{-4pt}
    \begin{minipage}{0.475\columnwidth}
        \caption*{Original}\vspace{-10pt}
        \begin{tikzpicture}
		    \node[anchor=south west,inner sep=1] (img) at (0,0) {\includegraphics[width=\columnwidth]{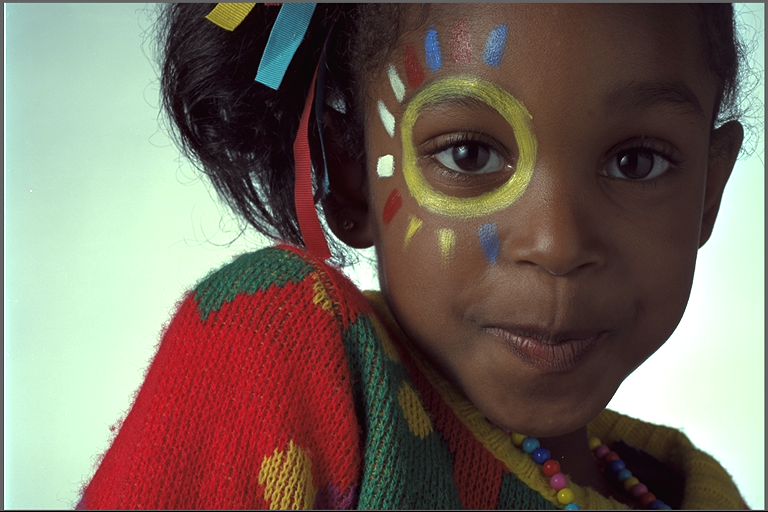}}; 
		    \begin{scope}[x={(img.south east)},y={(img.north west)}]
		    	\draw[green,thick] (0.5404,0.7832) rectangle (0.7747,0.5488);
		    \end{scope}  
	    \end{tikzpicture}
        \caption*{BPP / PSNR}\vspace{-3pt}
        \label{fig:kdm15_org}
    \end{minipage}
    \begin{minipage}{0.475\columnwidth}
        \caption*{BPG}\vspace{-10pt}
        \begin{tikzpicture}
		    \node[anchor=south west,inner sep=1] (img) at (0,0) {\includegraphics[width=\columnwidth]{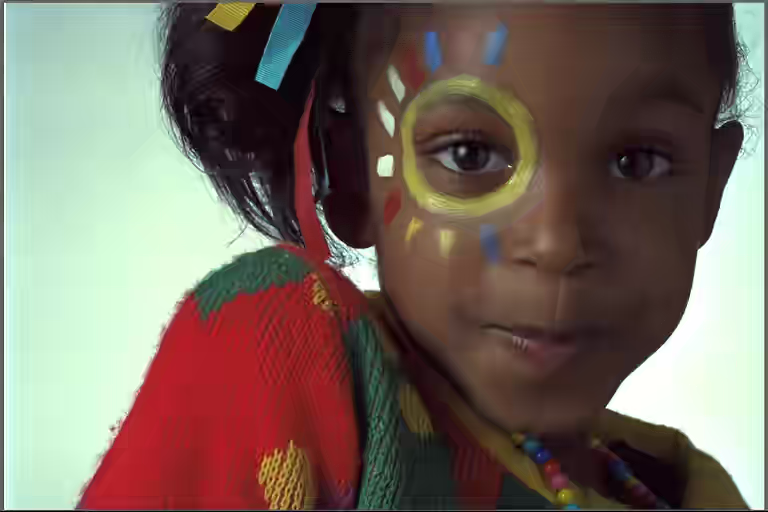}}; 
		    \begin{scope}[x={(img.south east)},y={(img.north west)}]
		    	\draw[green,thick] (0.5404,0.7832) rectangle (0.7747,0.5488);
		    \end{scope}  
	    \end{tikzpicture}
        \caption*{0.101 / 30.34}\vspace{-3pt}
        \label{fig:kdm15_bpg}
    \end{minipage}
    \begin{minipage}{0.475\columnwidth}
        \caption*{GDN-Joint\cite{minnen_joint}}\vspace{-10pt}
        \begin{tikzpicture}
		    \node[anchor=south west,inner sep=1] (img) at (0,0) {\includegraphics[width=\columnwidth]{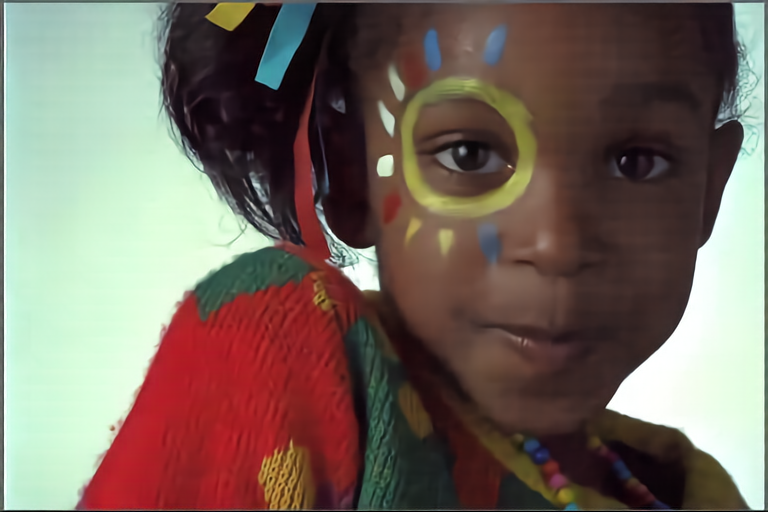}}; 
		    \begin{scope}[x={(img.south east)},y={(img.north west)}]
		    	\draw[green,thick] (0.5404,0.7832) rectangle (0.7747,0.5488);
		    \end{scope}  
	    \end{tikzpicture}
        \caption*{0.102 / 30.62}\vspace{-3pt}
        \label{fig:kdm15_joint}
    \end{minipage}
    \begin{minipage}{0.475\columnwidth}
        \caption*{SelfVAE-Joint}\vspace{-10pt}
        \begin{tikzpicture}
		    \node[anchor=south west,inner sep=1] (img) at (0,0) {\includegraphics[width=\columnwidth]{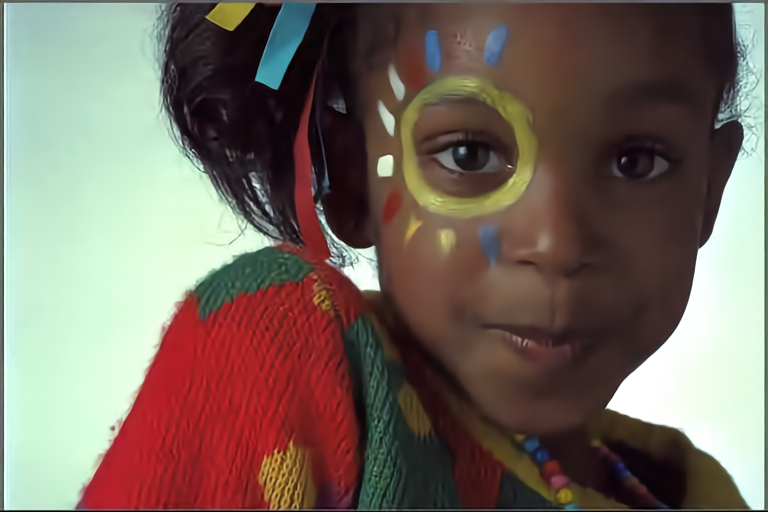}}; 
		    \begin{scope}[x={(img.south east)},y={(img.north west)}]
		    	\draw[green,thick] (0.5404,0.7832) rectangle (0.7747,0.5488);
		    \end{scope}  
	    \end{tikzpicture}
        \caption*{0.102 / 30.79}\vspace{-3pt}
        \label{fig:kdm15_selfonn}
    \end{minipage}\\
    \begin{minipage}{0.475\columnwidth}
        \includegraphics[width=\columnwidth]{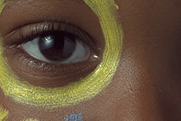}
        \label{fig:kdm15_org}
    \end{minipage}
    \begin{minipage}{0.475\columnwidth}
        \includegraphics[width=\columnwidth]{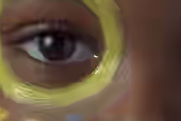}
        \label{fig:kdm15_org}
    \end{minipage}
    \begin{minipage}{0.475\columnwidth}
        \includegraphics[width=\columnwidth]{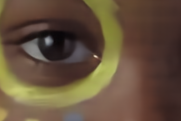}
        \label{fig:kdm15_org}
    \end{minipage}
    \begin{minipage}{0.475\columnwidth}
        \includegraphics[width=\columnwidth]{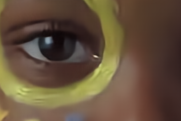}
        \label{fig:kdm15_org}
    \end{minipage}  \vspace{-12pt}
    \caption{Visual evaluation on KODIM 14 and KODIM 15 from the KODAK dataset at approximately equal bitrates.}
    \label{fig:visual}
\end{figure*}

\subsection{Experimental Results}
We compare the performance of the Self-VAE codec with those of GDN-based codec~\cite{minnen_joint} and benchmark BPG codec. Figure~\ref{fig:rd_curve} shows rate-distortion curve on Kodak test set. The curve is generated on low to mid bit-rate range. In all bit-rate regions on the curve, Self-VAE network gives superior results compared to ~\cite{minnen_joint}. BD-rate~\cite{bdrate} improvements provided by the proposed codec over~\cite{minnen_joint} and BPG are presented in Figure~\ref{fig:rate_savings}. Our Self-VAE model saves \% 7.78 over the anchor model~\cite{minnen_joint}. 

For qualitative visual comparison, Figure~\ref{fig:visual} shows visual quality of images obtained by different algorithms at nearly the same bit-rate. Our Self-VAE codec surpasses both the BPG codec and GDN-based codec~\cite{minnen_joint} in terms of PSNR and visual quality. If cropped sections are inspected closely, BPG has some serious compression artefacts, whereas the GDN-based model produces noticeably blurry images that lack texture details. On the other hand, our Self-VAE model preserves more texture details while avoiding blocking artefacts.

The LPIPS scores  \cite{lpips2018} (lower is better) tabulated in Table~\ref{lpips} verifies that the perceptual quality of images generated by the Self-VAE codec is superior to those of GDN-based codec~\cite{minnen_joint} and BPG.

In terms of model complexity, the Self-VAE model has $q = 3$ times more learnable parameters. However, this does not affect the inference time since $q$ convolutions in each generative neuron are performed in parallel.


\section{Conclusion}
\label{conc}
\vspace{-7.5pt}
We propose a novel Self-VAE network based on SOLs that are composed of generative neurons to replace VAE architectures that are based on convolutional neurons. Experimental results demonstrate that when widely used convolution and GDN layers in the state-of-the-art learned image codecs are replaced by SOL, the RD performance and visual quality improves significantly compared to their anchor models using the same entropy/context model. 
Since the proposed SOL is a versatile network layer, it can be directly plugged into any learned image codec to achieve superior performance.


\clearpage
\bibliography{references}
\bibliographystyle{IEEEtran}
\end{document}